\pdfoutput=1
\documentclass[%
 reprint,
 amsmath,amssymb,
 aps,
]{revtex4-2}

\usepackage{graphicx}
\usepackage{dcolumn}
\usepackage{bm}
\usepackage{physics}
\usepackage{dsfont}

\usepackage{xcolor} 
\usepackage{soul} 

\usepackage[colorlinks=true,linkcolor=blue,citecolor=blue,urlcolor=blue]{hyperref}
\usepackage{placeins} 

\begin{document}


\title{Eigen-SNAP gate for photonic qubits in a cavity-transmon system}


\author{Marcus Meschede,$^1$}
\email{marcus.meschede-2@uni-hamburg.de}

\author{Ludwig Mathey$^{1,2}$}
\affiliation{
$^1$Center for Optical Quantum Technologies and Institute for Quantum Physics, University of Hamburg,
22761 Hamburg, Germany\\
$^2$The Hamburg Center for Ultrafast Imaging, 22761 Hamburg, Germany\\
}


\date{June 30, 2025}

\begin{abstract}
In the pursuit of robust quantum computing, we put forth a platform based on photonic qubits in a circuit-QED environment. Specifically, we propose a versatile two-qubit gate based on two cavities coupled via a transmon, constituting a selective number-dependent phase gate operating on the in-phase eigenmodes of the two cavities, the Eigen-SNAP gate. This gate natively operates in the dispersive coupling regime of the cavities and the transmon, and operates by driving the transmon externally, to imprint desired phases on the number states. As an example for the utility of the Eigen-SNAP gate, we implement a $\sqrt{\text{SWAP}}$ gate on a system of two logical bosonic qubits encoded in the cavities. Further, we use numerical optimization to determine the optimal implementation of the $\sqrt{\text{SWAP}}$. We find that the fidelities of these optimal protocols are only limited by the coherence times of the system's components. These findings pave the way to continuous variable quantum computing in cavity-transmon systems.

\end{abstract}

\maketitle


\section{\label{sec:Introduction}Introduction\protect\\ }
\begin{figure}[t]
    \centering
    \includegraphics[width=\columnwidth]{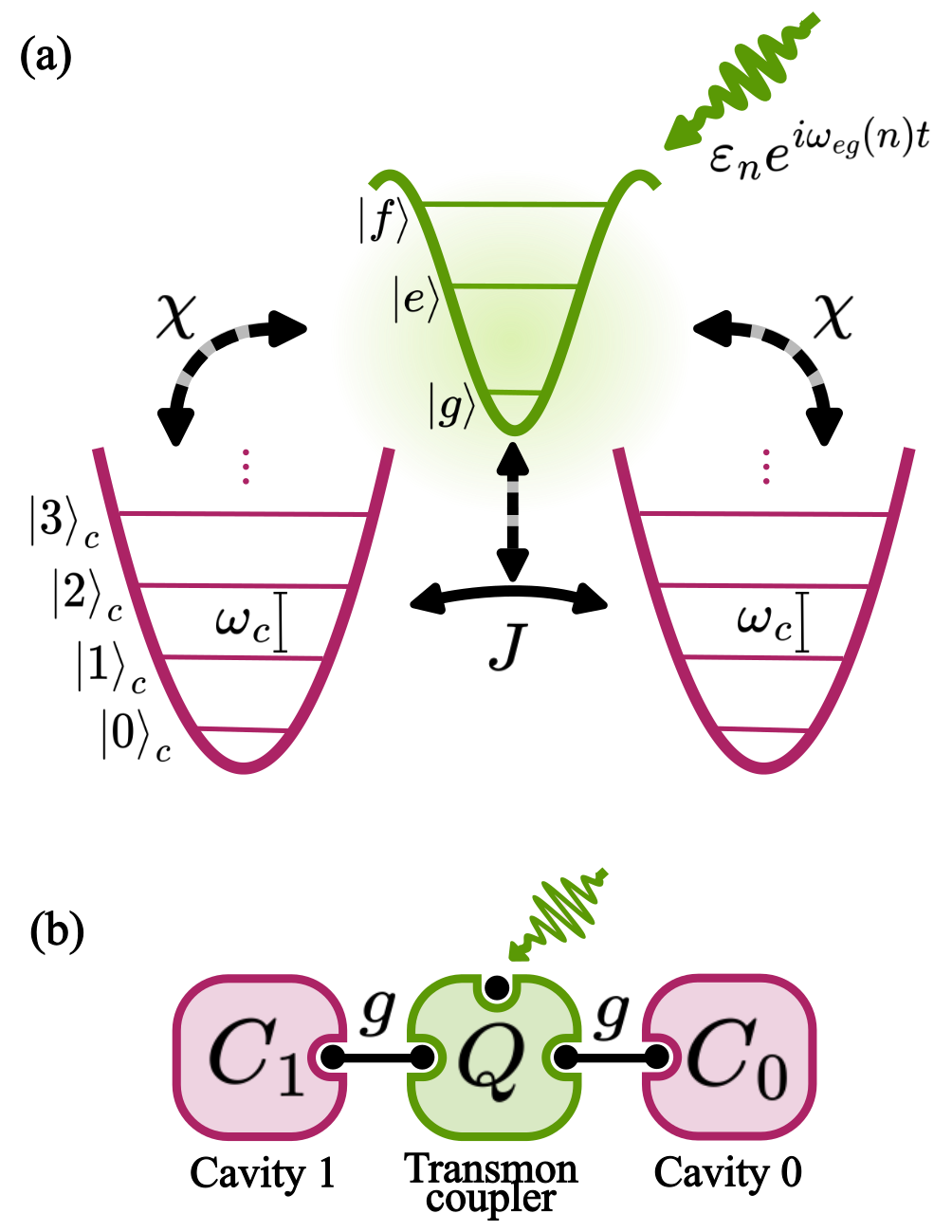}
    \caption{(a) We consider two cavities with the mode frequency $\omega_c$, coupled via a transmon. In the dispersive limit, the cavity modes interact with the transmon via a cross-Kerr nonlinearity, and with each other, via tunneling process with a tunneling constant, for which the sign is controlled by the transmon state. (b) The Hamiltonian Eq.~\eqref{equ:TwoCavDispHam} is the dispersive limit of two cavities coupled to a single transmon with a Jaynes-Cummings-like interaction.}
    \label{fig:Setup}
\end{figure}
High-fidelity quantum operations have been demonstrated on various qubit platforms, including superconducting qubits \cite{nakamuraCoherentControlMacroscopic1999, blaisCircuitQuantumElectrodynamics2021}, trapped ions \cite{ciracQuantumComputationsCold1995, blattQuantumSimulationsTrapped2012}, and neutral atoms \cite{jakschFastQuantumGates2000, adamsRydbergAtomQuantum2020}.
Despite significant progress in mitigating decoherence processes, the fidelities necessary for practical and scalable quantum computing might not be achievable without fault-tolerant hardware \cite{campbellRoadsFaulttolerantUniversal2017}. 
However, fault-tolerant approaches usually have a substantial resource overhead \cite{fowlerSurfaceCodesPractical2012}.

Photonic qubits are potentially a platform that can incorporate error-correctable properties by construction, as logical qubits are encoded in the much larger Hilbert spaces of cavity modes \cite{joshiQuantumInformationProcessing2021a, grimsmoQuantumComputingRotationSymmetric2020,gottesmanEncodingQubitOscillator2001, albertPerformanceStructureSinglemode2018}. 
Moreover, being implemented in cavity modes, photonic qubits exhibit long coherence times  \cite{reagorQuantumMemoryMillisecond2016}. 
Universal control over a single photonic mode is gained by coupling an ancilla qubit, such as a transmon, to the cavity and applying both cavity displacements and ancilla drives \cite{maQuantumControlBosonic2021}.
With these controls, logical operations on a single qubit can be constructed via numerical optimizations schemes, e.g. GRAPE \cite{heeresImplementingUniversalGate2017}. 
Alternatively, operations can be built from well-understood subroutines in a modular fashion, such as the selective number-dependent arbitrary phase gate (SNAP) \cite{krastanovUniversalControlOscillator2015, eickbuschFastUniversalControl2022}. 

Two-qubit gates are significantly harder to implement because a generic coupling between two photonic modes can drive the qubit state out of the logical subspace \cite{hongMeasurementSubpicosecondTime1987}. 
Therefore, the design of a possible two-qubit gate must be tailored to the specific coupling mechanism. 
Previous work has explored two-qubit entangling gates via bilinear coupling \cite{gaoProgrammableInterferenceTwo2018,  gaoEntanglementBosonicModes2019, zhangEngineeringBilinearMode2019, tsunodaErrorDetectableBosonicEntangling2023}, via dispersive coupling between the two cavities and a single ancilla \cite{rosenblumCNOTGateMultiphoton2018}, and a combination of dispersive coupling with a SNAP gate \cite{xuDemonstrationControlledPhaseGates2020}.

In this work, we propose to use a transmon state dependent bilinear coupling in combination with a local transmon drive, to design a continuous family of entangling gates. 
This specific coupling term is naturally motivated by the dispersive Hamiltonian of two photonic modes coupled to a single transmon, as sketched in Fig.~\ref{fig:Setup}.
We first design a selective number-dependent phase gate acting on the eigenmodes of the two-qubit system, which we call Eigen-SNAP gate. The main physical process is depicted as a level-diagram in Fig.~\ref{fig:S2_SWAP}~(a). The frequency difference of the first excited and the ground state of the transmon, $\omega_{eg}$, is shifted via the dispersive coupling to the in-phase superposition of the cavity modes, with the number states $n$, as discussed below. We propose to drive the system near the resulting transition frequencies $\omega_{eg}(n)$. The frequency shift due to the dispersive coupling supports imprinting a phase in a way that is state-selective on the number state $n$, which is the key mechanism for the Eigen-SNAP gate. We then use the Eigen-SNAP gate to construct the entangling gate for a range of error-correctable photonic qubit encodings. 
To further improve on this modularly constructed entangling gate, we numerically optimize its parametrization. This results in higher fidelity gates at shorter gate times. 
In Section \ref{sec:Model}, we describe the system Hamiltonian and the driving scheme. We introduce the Eigen-SNAP gate in Section \ref{sec:Method} and show how it can be used to implement an entangling gate on photonic qubits of various logical encodings. Going beyond the analytical entangling gate, in Section \ref{sec:Results} we evaluate its fidelity for finite gate times and for an optimal implementation of the gate via numerical optimization.

\section{\label{sec:Model}Cavity-transmon system\protect\\ }
We consider a system that consists of two cavities with frequencies $\omega_{c,\tau}$, $\tau = 0,1$, each dispersively coupled to a common transmon qubit with energy levels $\omega_j$, where $j =g,e,f$.
Additionally, the two cavities are coupled via a bilinear interaction that depends on the transmon state, see Fig.~\ref{fig:Setup}~(a). We illustrate our method by truncating the ancilla qubit to the lowest two energy levels $\ket{g}$ and $\ket{e}$. The Hamiltonian of the truncated system is
\begin{equation}
H_\text{disp} = H_0 + H_\chi + H_J
\label{equ:TwoCavDispHam}
\end{equation}
with the terms
\begingroup
\renewcommand{\arraystretch}{2.0} 
\begin{equation}
\begin{aligned}
H_0 &= \omega_g \sigma^{gg} + \tilde{\omega}_e \sigma^{ee}+\sum_{\tau=0,1} \omega_{c,\tau}\;a_\tau^\dagger a_\tau \\
H_\chi &= -\sum_{\tau=0,1}   \;\chi^{ge}_\tau\sigma^{z}_{eg}a_\tau^\dagger a_\tau \\
H_J &= -J_{01}^{ge}\sigma^{z}_{eg}\left(a_1^\dagger a_{0} + a_{0}^\dagger a_1 \right) \quad . \\
\end{aligned}
\label{equ:TwoCavDispHamParts}
\end{equation}
\endgroup
The operators $\sigma^{jk} = |j\rangle\langle k|$ and $\sigma^{z}_{jk} = |j\rangle\langle j|-|k\rangle\langle k|$ act on the transmon states, while the bosonic operators $a_\tau$ act on the cavities.
The term $H_0$ describes the bare energies of the cavities and transmon. The dispersive coupling of strength $\chi^{ge}_\tau$ between the transmon and the cavities, and the transmon state dependent bilinear coupling of strength $J_{01}^{ge}$ are represented by $H_\chi$ and $H_J$, respectively.

We derive the Hamiltonian $H_\text{disp}$ in Appendix~\ref{app:SW} as the effective Hamiltonian of a system of two cavities with a Jaynes-Cummings-like coupling to a single few-level transmon, see Fig.~\ref{fig:Setup}(b).
In the regime of weak coupling $g_\tau$ and large detuning between the cavities and transmon qubit energies, $g_\tau \ll \abs{\omega_{c,\tau}-(\omega_j - \omega_k)} $, we obtain the effective Hamiltonian in Eq.~\eqref{equ:TwoCavDispHam}, with the coefficients $\chi^{jk}_\tau = g_\tau^2/\Delta_\tau^{jk}$, $\Delta_\tau^{jk} = \omega_\tau + \omega_j - \omega_k$ , $\tilde{\omega}_e = \omega_e - \sum_\tau\chi^{ge}_\tau$, and 
\begingroup
\renewcommand{\arraystretch}{1.5} 
\begin{equation}
    \begin{aligned}
    J_{01}^{jk} &= \frac{1}{2}g_{0}g_{1}\left(\frac{1}{\omega_{c,0}+\omega_j-\omega_k}+\frac{1}{\omega_{c,1}+\omega_j-\omega_k}\right) \; .
    \end{aligned}
\end{equation}
\endgroup

To construct an entangling gate, we tune the cavities on resonance with each other and assume the same coupling strength to the transmon,  $\omega_c = \omega_{c,\tau} $ and $g = g_\tau$. Consequently, the dispersive coupling strengths reduce to a single value, and we define $\chi = \chi_\tau^\text{ge}$. However, the cavities remain detuned from the transmon energy level spacings.

In this regime, the dispersive Hamiltonian in Eq.~\eqref{equ:TwoCavDispHam} can be written in diagonal form, see Appendix~\ref{app:Diag}, 
\begin{equation}
\begin{aligned}
H_\text{disp} =&\; \left(\frac{1}{2}\omega_{eg}-2\chi b^\dagger_+b_+\right)\sigma^{z}_{eg}\\
&+\omega_c\left(b^\dagger_+b_+ +b^\dagger_-b_-  \right) \\
\end{aligned}
\label{equ:TwoCavDispHamEigen}
\end{equation}

with $\omega_{eg} = \tilde{\omega}_e-\omega_g$ and the eigenmode operators
\begin{equation}
    \begin{aligned}
b_\pm&= \frac{1}{\sqrt{2}}(a_1 \pm a_0) .
\end{aligned}
\label{equ:eigenmodes}
\end{equation} 
We write a general number state of these modes as
\begin{equation}
    \ket{n}_+\ket{m}_- =\frac{1}{\sqrt{n!m!}}\left(b_+^\dagger\right)^n\left(b_-^\dagger\right)^m\ket{0}_+\ket{0}_- \quad .
\label{equ:EigenModeOccupationNumberState}
\end{equation}
We note that the accumulated relative phase of the time evolution of state $\ket{g}\ket{n}_+\ket{m}_-$ and $\ket{e}\ket{n}_+\ket{m}_-$ under the interaction term $2\chi \sigma_{eg}^z b_+^\dagger b_+$ differs by a multiple of $2\pi$ at integer multiples of the time $T=\pi/\chi$. In the design of two-qubit gates between the cavity modes, we will use multiples of this time $T$ for the gate time $T_g$.

In addition to the static dispersive Hamiltonian \eqref{equ:TwoCavDispHam}, we apply a local drive on the transmon qubit
\begin{equation}
\begin{aligned}
        H_D &= \varepsilon(t)\sigma^{ge}
        + h.c. \\
\end{aligned} 
\label{equ:DrivingQubitHamiltonians}
\end{equation}
which we parametrize as
\begin{equation}
\begin{aligned}
        \varepsilon(t) = \sum_n \varepsilon_n e^{i\omega_{eg}(n) t} e^{i\Theta_n(t)}
\end{aligned}
\label{equ:DrivingQubitParametrization}
\end{equation}
with a time dependent phase $\Theta_n(t)$ and the driving frequencies $\omega_{eg}(n) = \omega_{eg}-4\chi n $, as depicted in Fig.~\ref{fig:S2_SWAP}~(a). In the optimization, we allow for a frequency modulation $\Delta \omega_n$.

\section{\label{sec:Method}Realizing entangling gates\protect\\ }
In this section, we introduce the eigenmode selective occupation number-dependent gate (Eigen-SNAP). Further, we use the Eigen-SNAP gate to implement a family of entangling gates on photonic qubits. 

\subsection{Eigen-SNAP gate}
\label{subsec:eigen_snap_gate}
To construct the Eigen-SNAP gate, we drive the transmon mode only. As shown in Eq.~\eqref{equ:TwoCavDispHamEigen}, the transmon energy is modified by the $n_+$ state. Therefore, the transmon drive of Eq.~\eqref{equ:DrivingQubitHamiltonians} can be used to apply different phases $\theta$ onto individual occupation number states $\ket{n}_+$ of the $b_+$ eigenmode. Specifically, the action of the Eigen-SNAP gate is

\begin{equation}
S^+_n(\theta) = e^{i\theta \ketbra{n}_+} 
\label{equ:SNAPEigenGate}
\end{equation}
where for the driving amplitude we assume $\varepsilon_m=0$ for $m\neq n$. Here, the operator $\ketbra{n}_+$ projects onto the eigenmode number states of the $b_+$ mode. Similar to the SNAP gate in the single oscillator-ancilla system, the $S^+_n$ relies on the accumulation of the geometric phase. The geometric phase is realized by two consecutive $\pi$-pulses with a relative phase offset in the transmon subspace of states $\ket{g}$ and $\ket{e}\ket{n}_+\ket{m}_-$.

\begin{figure*}[t]
    \centering
    \includegraphics[width=\textwidth]{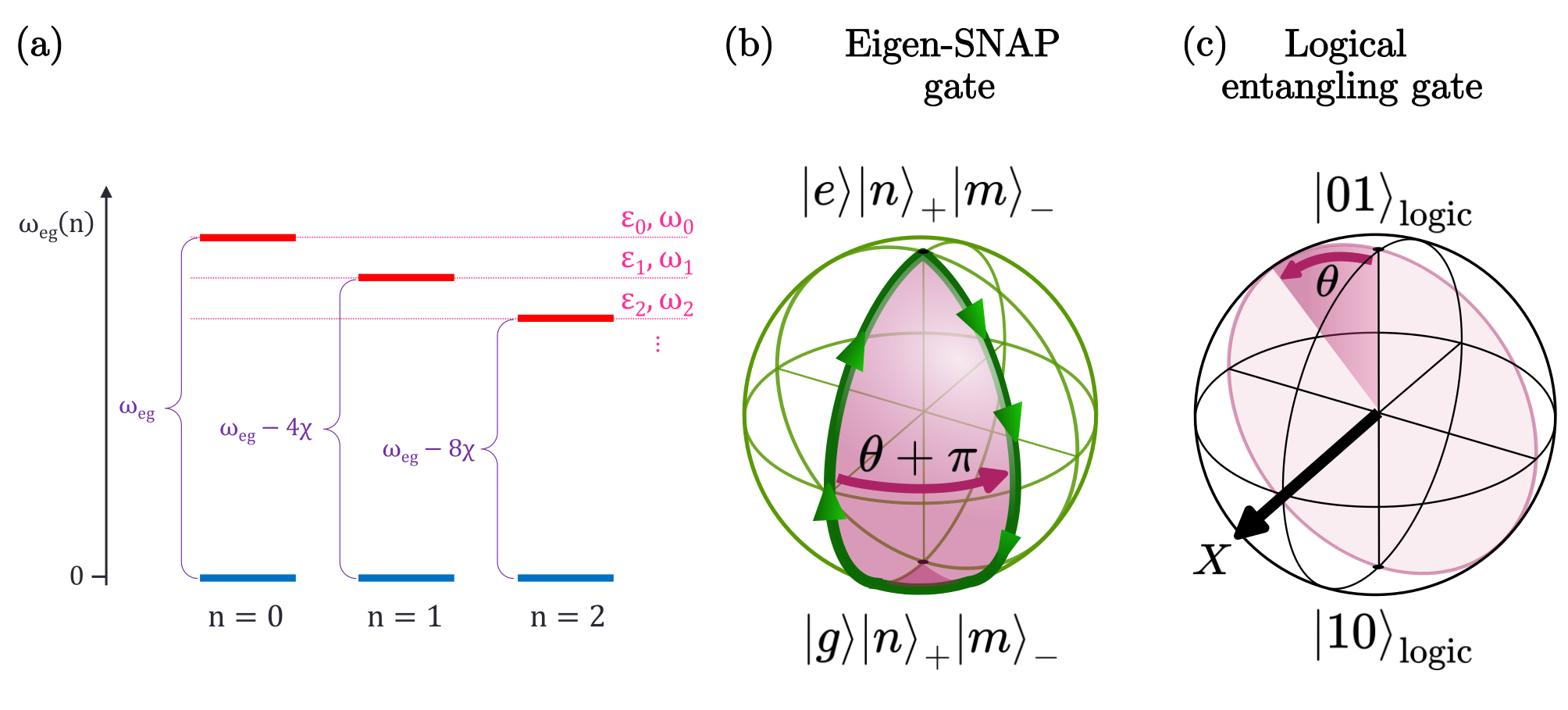}
    \caption{(a) Level structure of a dispersively coupled cavity-transmon system, as in Eq. (4). The frequency of the transmon $\omega_{eg}$ is detuned by the in-phase eigenmode occupation number. (b) The Eigen-SNAP gate $S^+_n(\theta)$ (Eq.~\eqref{equ:SNAPEigenGate}) is realized by two consecutive $\pi$-pulses in the space spanned by $\ket{g}\ket{n}_+\ket{m}_-$ and $\ket{e}\ket{n}_+\ket{m}_-$. The first $\pi$-pulse has the phase zero, while the second $\pi$-pulse has the phase $\pi+\theta$. This results in a geometric phase $\theta$ that is imprinted on the number state $\ket{n}_+$. (c) We propose to construct an entangling gate by applying the Eigen-SNAP gate with the same phase to all odd number states. In the logical encodings of the form Eq.~\eqref{equ:EvenOdd} this corresponds to a rotation around the X-axis in the $\ket{01}_{\text{logic}} / \ket{10}_{\text{logic}}$ logical subspace. This provides a gate for Binomial, even-legged cat or single-rail encodings.}
    \label{fig:S2_SWAP}
\end{figure*}

To implement this trajectory we use the parametrization of the driving term in Eq.~\eqref{equ:DrivingQubitParametrization} such that the amplitude $\varepsilon_n$ fulfills $\varepsilon_n T_g  = \pi$, while setting the gate times $T_g$ to a multiple of $T = \pi/\chi$. At these stroboscopic times, the system does not acquire an additional phase due to the dynamic phase evolution. At gate times $T_g$, that are not integer multiples of $T$, the amplitudes $\varepsilon_n$ must be chosen individually to compensate the phase. To generate a geometric phase, the time-dependent phase $\Theta_n(t)$ of Eq.~\eqref{equ:DrivingQubitParametrization} is given by

\begin{equation}
\Theta_n(t) = 
\begin{cases} 
0 & \text{if} \quad t < T_g/2 \\ 
\theta + \pi & \text{if} \quad t \geq T_g/2   \;.
\end{cases}
\label{equ:PulseProtocol}
\end{equation} 
The resulting state trajectory in the rotating frame of the eigenmode is illustrated in  Fig.~\ref{fig:S2_SWAP}(a). For the limit of $T_g\chi\gg 1$, and hence a small driving amplitude $\varepsilon_n$, the correspondence between this pulse sequence and the $S^+_n$ gate becomes exact, see App.~\ref{app:SNAPGate}. We therefore call it the adiabatic parametrization.
Finally, we note that only the number states of the $b_+$ mode can be addressed individually since they are energetically separated via the dispersive coupling to the transmon.
Multiple $S^+_n(\theta)$ gate pulses can be applied simultaneously on different occupation number states $\ket{n}_+$ by overlaying the corresponding pulses as alluded to in the summation in Eq.~\eqref{equ:DrivingQubitParametrization}. However, for finite times this introduces significant errors that we correct through the numerical optimization in Sec.~\ref{sec:Results}. \\
We note that for implementation of the Eigen-SNAP gate $S^+_n(\theta)$ in a three-level system, we modify the driving frequency corresponding to the individual number states. Explicitly, in the driving parametrization (Eq.~\eqref{equ:DrivingQubitParametrization}) we substitute  $4\chi^{ge} \rightarrow 2\left(2\chi^{ge} - \chi^{ef}\right)$, where $\chi^{ef}$ is the dispersive coupling to the $\ket{e}-\ket{f}$ subspace, see Appendix~\ref{app:SNAPGate}. 

\subsection{Logical entangling gate}
To construct an entangling gate, we translate the phase on the number states into a logical operation on the cavity modes.
We propose to use an entangling gate that imprints a phase $\theta$ on each odd number state, i.e.:
\begin{equation}
S_\text{odd}^+(\theta) = \prod_{n\in odd}^{N_{\text{max}}} S^+_n(\theta) \quad 
\label{equ:Soddgate}
\end{equation}
where the largest eigenmode occupation number is given by  
\begin{equation}
    N_{\text{max}} = \max\limits_{\ket{p,q}_c \in \ket{\cdot,\cdot}_\text{logic}} (p+q)  \quad .
\end{equation}
Hence, the $S_\text{odd}^+(\theta)$ transformation maps states $\ket{n}_+$ with an odd number $n$ onto $\exp{i \theta}\;\ket{n}_+$ and all states $\ket{n}_+$ with an even number $n$ onto $\ket{n}_+$. Therefore, $S_\text{odd}$ acts only on the subspace of the in-phase degree of freedom $b_+$, spanned by the states $\ket{n}_+$ , see Eqs.~\eqref{equ:eigenmodes} and \eqref{equ:EigenModeOccupationNumberState}, but leaves the subspace of the out-of phase degree of freedom $b_-$ invariant.
We propose to use this gate for encodings for which each codeword $\ket{0/1}_\text{logic}$ consists of either only even or only odd number states. We adopt the notation that if both logical states are composed of the same parity Fock states, we write $\ket{0/1}_\text{logic}$, and if they are composed of opposite parity Fock states $\widetilde{\ket{0/1}}_\text{logic}$. We require that the two qubits must be implemented with the same encoding. 
These requirements include error-correctable encodings like the Binomial codes \cite{michaelNewClassQuantum2016} and cat codes \cite{cochraneMacroscopicallyDistinctQuantumsuperposition1999} with an even number of legs. They also apply to non-error correctable encodings such as single-rail encoding in the lowest two Fock states $\ket{0/1}_{\text{logic}} = \ket{0/1}_c$. 

The logical operation induced by the gate $S^+_\text{odd}$ mixes the logical states $\ket{01}_\text{logic}$ and $\ket{10}_\text{logic}$  ($\widetilde{\ket{01}}_\text{logic}/\widetilde{\ket{10}}_\text{logic}$). In this subspace, it corresponds to an X-axis rotation, see Fig.~\ref{fig:S2_SWAP}(b), summarized in Table~\ref{tab:parity-expressions}. The logical states $\ket{00}_\text{logic}$ and $\ket{11}_\text{logic}$ are unaffected. The gate in the same parity case is commonly called $\text{SWAP}^\alpha$, with $\alpha = 2\theta/\pi$. Apart from the above requirements, the $S_\text{odd}^+(\theta)$ entangling gate is agnostic to the specific codewords in use. 
\begin{table}[htbp]
    \centering
    \renewcommand{\arraystretch}{2.0} 
    \begin{tabular}{p{0.25\columnwidth} p{0.25\columnwidth} p{0.4\columnwidth}} 
        \hline\hline 
        \multicolumn{2}{c}{type of encoding} & $S_\text{odd}^+(\theta)$ \\ 
        \hline 
         $\ket{0/1}_{\text{logic}}$ & Eq.~\eqref{equ:EvenOdd} & $\exp\left(i\theta/2(\mathds{1}-\sigma^x)\right)$ \\ 
        $\widetilde{\ket{0/1}}_{\text{logic}}$ & Eq.~\eqref{equ:EvenOdd_OppositeParity} & $\exp\left(i\theta/2 (\mathds{1}+\sigma^x)\right)$ \\ 
        \hline\hline 
    \end{tabular}
    \caption{Logical operation in the $\ket{01}_\text{logic}/\ket{10}_\text{logic}$ subspace using the $S_\text{odd}^+(\theta)$ gate. The logical operation depends on the relative even/odd parity of the logical single qubit codewords.}
    \label{tab:parity-expressions}
\end{table}
The process is best illustrated by writing the two qubit logical basis in the eigenmode basis $\ket{n}_+\ket{m}_- $, as defined in Eq.~\eqref{equ:EigenModeOccupationNumberState} of the two cavity system. Further, we group them into all contributions with only even (odd) $\ket{\text{even}_+} $( $\ket{\text{odd}_+}$ )  number occupation states $\ket{n}_+$ of the $b^+$ mode. Then $\ket{00}_\text{logic}$ and $\ket{11}_\text{logic}$ only consist of even eigenmode occupation numbers. The $\ket{01}_\text{logic}$ and $\ket{10}_\text{logic}$ states additionally have odd number state contributions. While for both $\ket{01}_\text{logic}$ and $\ket{10}_\text{logic}$ the contribution is of equal magnitude, the two logical states differ in the relative phase of $\ket{\text{even}}$ and $\ket{\text{odd}}$. Hence, applying the same phase onto all even or odd contributions leads to a rotation in the $\ket{01}_\text{logic}$ and $\ket{10}_\text{logic}$ subspace. 

The reason for the slightly different logical entangling gate implemented by the operation $S^+_\text{odd}$ for the different cases of relative parity of the single cavity encoding lies in the decomposition into $\ket{\text{even}_+}$ and $\ket{\text{odd}_+}$ parts.

For single qubit logical encodings, in which the parity of the cavity Fock states of both $\ket{0}_{\text{logic}}$ and $\ket{1}_{\text{logic}}$ are the same
\begin{equation}
    \begin{aligned}
        \ket{01}_\text{logic} &= \ket{\text{even}_+} + \ket{\text{odd}_+}\\
        \ket{10}_\text{logic} &= \ket{\text{even}_+} - \ket{\text{odd}_+} \quad \quad 
    \end{aligned}
\label{equ:EvenOdd}
\end{equation}
while for encodings in which the parity of the cavity Fock states of both $\widetilde{\ket{0}}_{\text{logic}}$ and $\widetilde{\ket{1}}_{\text{logic}}$ are the opposite

\begin{equation}
    \begin{aligned}
        \widetilde{\ket{01}}_\text{logic} &= \ket{\text{even}_+} + \ket{\text{odd}_+}\\
        \widetilde{\ket{10}}_\text{logic} &= -\ket{\text{even}_+} + \ket{\text{odd}_+} \quad \quad .
    \end{aligned}
\label{equ:EvenOdd_OppositeParity}
\end{equation}
Then it follows that imprinting a phase on the odd eigenmode number states leads to the result in Table~\ref{tab:parity-expressions}.
\begin{figure*}[t]
    \centering
    \includegraphics[width=\textwidth]{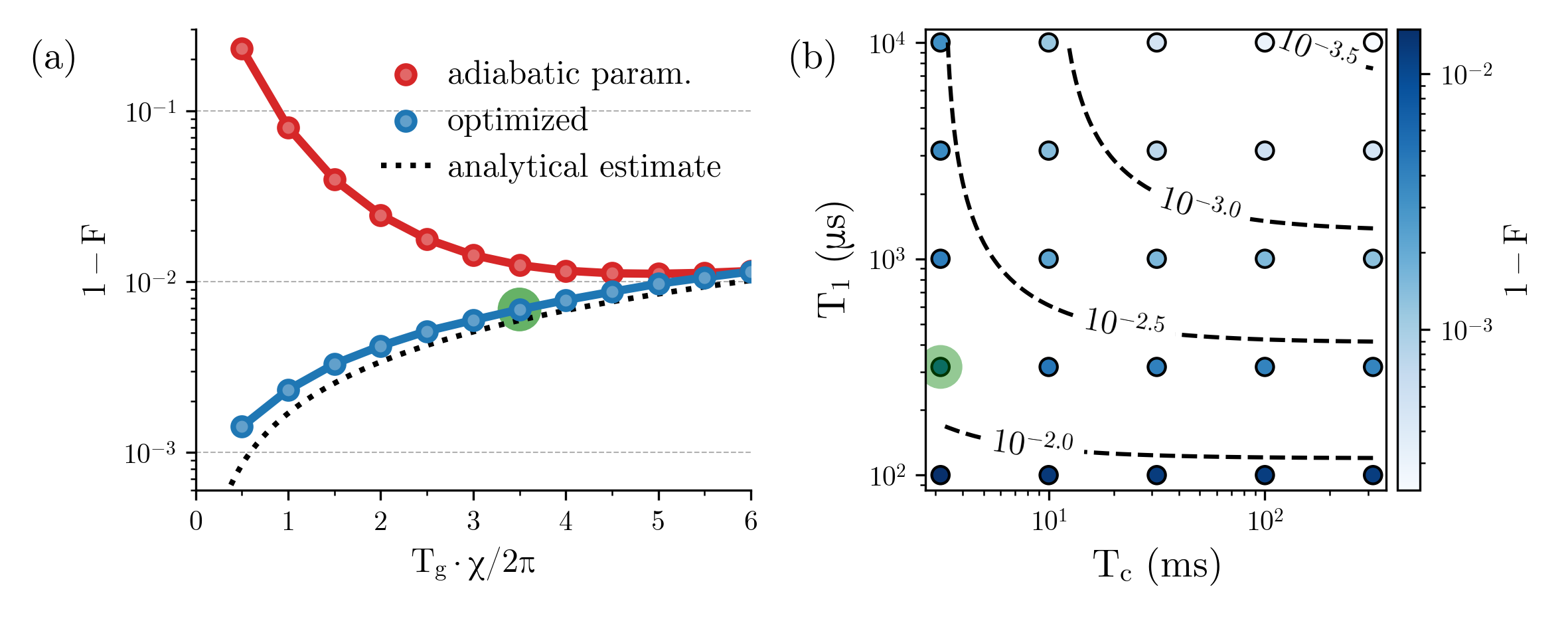}
    \caption{(a) Infidelity, as defined in Appendix~\ref{app:Fidelity}, of the $S^+_\text{odd}(\pi/2)$  gate on the Binomial encoding in Eq.~\eqref{equ:BinomCode} for different total gate times $T_g$. The fidelity with the adiabatic parametrization (red) of Eq.~\eqref{equ:Soddgate} is improved for larger gate times up to the analytical estimate of the minimum (dotted), see Eq.~\eqref{equ:FidTheoryMin}, set by incoherent errors of the system. The fidelity of the optimized parametrization (blue) is near the analytical estimate for all gate times. Here, $T_\text{loss} = 3.16 \,ms$, $T_1 = 316\, \mu s$, $T_2 = 1/2 T_1$.
    (b) Infidelity as a function of coherence times of the cavity $T_c$ and the transmon $T_1$, with $T_2 = 1/2 T_1$ for gate time $T_g\cdot \chi/2\pi=3.5$. The common data point to panel (a) is marked in green. The contour lines indicate the levels of the infidelity of the optimized pulses determined by fitting the coefficients of different error contributions, see Eq.~\eqref{equ:FidTheoryMin}, to the fidelities found in the optimization (blue). }
    \label{fig:Results}
\end{figure*}

\section{\label{sec:Results}Optimal pulse parametrization\protect\\ }
We evaluate the fidelity of the entangling gate $S^+_\text{odd}$ based on our Eigen-SNAP gate on a Binomial encoding for different gate times and decoherence strengths of the transmon and the cavities. We optimize the parametrization of the entangling gate and obtain fidelities at the limit of incoherent losses of the system. 

In the evaluation, we focus on the $S^+_\text{odd}(\pi/2)$ gate, corresponding to the $\sqrt{\text{SWAP}} $ gate and a Binomial encoding with average occupation number $\langle n\rangle=3$:
\begin{equation}
    \begin{aligned}
\ket{0}_\text{binom} = \frac{1}{2}\left(\ket{0}+\sqrt{3}\ket{4}\right)\\
\ket{1}_\text{binom} = \frac{1}{2}\left(\sqrt{3}\ket{2}+\ket{6}\right)\\
\label{equ:BinomCode}
\end{aligned}
\end{equation}
This encoding allows to correct for a single cavity photon loss and a single cavity dephasing event. We do not include the error correction protocol itself in this study. There exist larger Binomial encodings protecting against arbitrary orders of these cavity loss processes. Our method is also applicable to these encodings. 

We model the decoherence mechanism via a Lindblad master equation (Eq.~\eqref{equ:LindbladMasterEquation}).
The dominant decoherence mechanism and their Lindblad operators are the photon loss in each cavity $L_{c,0/1} = \sqrt{\kappa} a_{0/1}$, the transmon dephasing $L_z = \sqrt{\gamma_z} \sigma^z_{eg}$ and the transmon decay $L_- =\sqrt{\gamma_-} \sigma^{ge}$, with the rates  $\kappa=1/T_c$, $\gamma_-=1/T_1$ and $\gamma_z=1/T_2-1/(2T_1)$. $T_c$ denotes the relaxation time of the cavities, while $T_1$ and $T_2$ are the decay and dephasing times of the transmon. The coherence times of current superconducting cavity-qubit setups are on the order of milliseconds for the cavity relaxation time and coherence times for the transmon of $T_1=100\mu s$ and $T_2=50\mu s$. However, through error detection or error correction protocols on the transmon during the Eigen-SNAP protocol, higher effective coherence times might be achievable \cite{rosenblumFaulttolerantDetectionQuantum2018, reinholdErrorcorrectedGatesEncoded2020, maPathIndependentQuantumGates2020, landgrafFastQuantumControl2023}. Much larger relaxation times have also been demonstrated for cavity-qubit setups \cite{milulSuperconductingCavityQubit2023}. Hence, we evaluate our optimized gate for a range of coherence times $T_c$ and $T_1$ of the system components, where we assume $T_1 = 2 T_2$ for simplicity. 

The fidelity is calculated by averaging over a complete set of generators for the logical two-qubit Hilbert space, see Appendix~\ref{app:Fidelity}.

In an experimental realization the fidelity of the implemented states can be evaluated by first reconstructing the density matrix with state tomographic methods. These methods mostly rely on repeated state preparations and displacements to sample the phase space distribution, such as the Wigner function \cite{lutterbachMethodDirectMeasurement1997, gertlerExperimentalRealizationCharacterization2023}. Two-mode state tomography has been shown in \cite{rosenblumCNOTGateMultiphoton2018, gertlerExperimentalRealizationCharacterization2023}.
An optimal sampling strategy of the phase space has been introduced in \cite{krisnandaDemonstratingEfficientRobust2025} , further reducing the experimental overhead of state tomography. Adding measurement ancilla qubits to the individual cavities in our setup would make it amendable to these methods.

To estimate the limits imposed on our Eigen-SNAP based entangling gate protocol due to incoherent errors, we calculate the individual first-order corrections of the density matrix evolution due to loss processes.  
We find the analytical minimal infidelity set by the ideal Eigen-SNAP trajectory for the above encoding under the above decoherence processes to be, see Appendix~\ref{app:Fidelity},

\begin{equation}
    (1-F)_\text{analy. estimate} = \left( 6 \kappa + \frac{11}{32}\gamma_z +\frac{9}{128}\gamma_-\right)T_g \; . 
    \label{equ:FidTheoryMin}
\end{equation}

The loss of fidelity is proportional to the total photon number $2\langle n \rangle$ of an error correctable encoding. By construction, the loss in the cavity is independent of the logical state in the two-cavity system. The loss due to transmon decoherences during the Eigen-SNAP gate pulse however is highly dependent on the logical state of the cavities. 
To minimize the total gate time $T_g$, we apply the driving pulses for all involved occupation number states at the same time, approximately realizing the $S^+_\text{odd}(\pi/2)$ gate after gate time $T_g$. The resulting infidelity is shown for different total gate times in Fig.~\ref{fig:Results}(a). For large gate times, for which the adiabatic parametrization of Sec.~\ref{subsec:eigen_snap_gate} approaches the analytically exact gate,  the infidelity reaches the estimated minimum (analy. min.) set by the incoherent losses of the components of the system.

Shorter gate times are desirable to minimize the effect of incoherent losses. Therefore, we resort to numerical optimization using Gradient Ascent Pulse Engineering (GRAPE algorithm) to optimize the parameters of the drive (Eq.~\ref{equ:DrivingQubitParametrization}) with the loss function being the fidelity.

From the parametrization we optimize the frequency detuning $\Delta \omega$ and the complex amplitude $\varepsilon_n$ for all $n\leq 13$, starting from the gate Ansatz. Including a drive that is resonant with the even occupation number states allows the system to find a gate implementation up to a global phase. The loss trajectories were found to be smoother if we additionally allowed for the total gate time to be optimized. 

The optimization procedure is simplified by the fact that the driving of the transmon does not lead to transitions between different eigenmode occupation numbers (Appendix~\ref{app:SNAPGate}). Hence, in the optimization we only need to simulate the evolution of a single state, $\ket{\psi_0} =\left(1/\sqrt{14}\right)  \sum_{n=0} \ket{n}_+\ket{0}_-$, and optimize for an increased overlap with $S^+_\text{odd}(\pi/2)\ket{\psi_0}$ up to a global phase. This procedure keeps the parametrization agnostic to the logical encoding and leads to an optimized parametrization that retains symmetric characteristics, see Fig.~\ref{fig:ThetaOptimized} in Appendix~\ref{app:Optimized}. 

To evaluate the optimized pulse parametrization, we calculate the fidelity of the pulse for different gate times and dissipation strengths using the binomial encoding of Eq.~\eqref{equ:BinomCode} using the full master equation. We determine the weight of the different error coefficients of Eq.~\eqref{equ:FidTheoryMin} found in the numerical optimization through a linear fit. We observe, that the resulting fidelity is close to the analytical minimum for all gate times (Fig.~\ref{fig:Results}~(a)). Tuning the loss magnitudes individually, we find the optimized infidelity to scale with
\begin{equation}
    (1-F)_\text{optimized} \approx \left( 6.02 \kappa + 0.37\gamma_z +0.11\gamma_-\right)T_g \quad.
    \label{equ:FidOptimized}
\end{equation}
The scaling with $\kappa T_g$ and $\gamma_z T_g$, the dominant loss channels, is in good agreement with the analytically determined minimal infidelity. We therefore conclude that the optimized pulses are optimal. 

We determine the minimal infidelity of the optimized pulses for the specific gate time of $T_g\chi/\pi = 3.5$, we evaluate the optimized fidelity of the $\sqrt{\text{SWAP}}$ gate for different loss rates in Fig.~\ref{fig:Results}(b). The contour lines mark error levels found through the fitted Eq.~\eqref{equ:FidTheoryMin}.

\section{\label{sec:Conclusion}Conclusion\protect\\ }

In conclusion, we have proposed a quantum gate for photonic qubits in a cavity-transmon system, the Eigen-SNAP gate. This gate operates in the dispersive limit of two degenerate cavities coupled via a transmon. Thus, number states of the in-phase superposition of the two cavity modes shift the transmon frequency, proportionally to the occupation number. The operation of the gate consists of driving the transmon with two sequential $\pi$-pulses at a large subset of these frequencies, corresponding to a subset of the number states. The geometric phase of the $\pi$-pulses is imprinted on the subset of number states. We have discussed the case of driving all frequencies corresponding to the odd number states of the in-phase mode, based on the choice of qubit encoding. We note that our gate supports a wide range of other encodings in the cavity modes, by driving the frequencies that correspond to the modes that constitute the logical qubits. For the case of encoding in the even and odd number states, we have demonstrated the $\sqrt{\text{SWAP}}$ gate, based on the Eigen-SNAP gate, as an example for a quantum gate. Utilizing optimal control via GRAPE, we have demonstrated an adjusted set of pulses, which significantly improves the fidelity of the $\sqrt{\text{SWAP}}$ gate, such that it is only limited by the intrinsic decoherence of the cavities and the transmon.

With this gate architecture, we put forth a viable and versatile approach towards photonic quantum computing in circuit-QED environments, and, more generally, in continuous-variable quantum information processing.

\section*{Acknowledgments}
The project is financed by ERDF of the European Union and by ’Fonds of the Hamburg Ministry of Science, Research, Equalities and Districts (BWFGB)’. We acknowledge funding by the Cluster of Excellence “Advanced Imaging of Matter” (EXC 2056) Project No.390715994.
\newpage

\bibliography{CavitySqSWAP1}

\appendix

\section{\label{app:SW}Dispersive limit derivation}
We perform a Schrieffer-Wolff transformation of a single three-level system coupled to multiple cavity modes:
\begin{equation}
  \begin{aligned}
H_0 &=  \omega_g \sigma^{gg} + \omega_e \sigma^{ee}+ \omega_f \sigma^{ff}+\sum_\tau\omega_{c,\tau}\;a_\tau^\dagger a_\tau \\
V &= \sum_\tau g_\tau a_\tau\left(\sigma^{eg}+\sigma^{fe}\right)+ h.c. \\
\end{aligned}
\label{equ:H0V}
\end{equation}
We have adopted the notation $\sigma^{nm} = \vert n\rangle \langle m \vert$. Also $\sigma^z_{nm} = \vert n\rangle \langle n \vert - \vert m\rangle \langle m \vert$. Assuming $g_\tau \ll \abs{\omega_{c,\tau}-(\omega_j - \omega_k)} $ and using the transformation $U = \exp{e^{-S}}$ with 

\begin{equation}
  \begin{aligned}
S =& \; -\sum_\tau g_\tau \bigg(a_\tau\left(\frac{1}{\Delta^{ge}_\tau}\sigma^{eg}+\frac{1}{\Delta^{ef}_\tau}\sigma^{fe} \right)\\
&- a_\tau^\dagger \left(\frac{1}{\Delta^{ge}_\tau}\sigma^{ge}+\frac{1}{\Delta^{ef}_\tau}\sigma^{ef} \right)\bigg)
\end{aligned}
\end{equation}
for which $[S,H_0]=-V$, the full Hamiltonian is given by

\begin{equation}
    \begin{aligned}
H &= H_0 + \frac{1}{2}[S, V]\\
&=\omega_g \sigma^{gg} + \tilde{\omega}_e \sigma^{ee}+ \tilde{\omega}_f \sigma^{ff}+\sum_{\tau}\omega_{c,j}\;a_{\tau}^\dagger a_{\tau} \\
&-\sum_{\tau} \bigg[  \;a_{\tau}^\dagger a_{\tau} \left(\chi^{ge}_{\tau}\sigma^{z}_{eg}+\chi^{ef}_{\tau}\sigma_{fe}^z\right) \\
&-\frac{1}{2}\left(\chi^{ge}_{\tau} -\chi^{ef}_{\tau}\right)\left(a_{\tau}a_{\tau}\sigma^{fg}+a^\dagger_{\tau}a^\dagger_{\tau}\sigma^{gf}\right)\bigg] \\
&-\sum_{\tau> \tau'} \bigg[ \;\left(a_{\tau}^\dagger a_{\tau'} + a_{\tau'}^\dagger a_{\tau} \right)\left(J^{ge}_{\tau \tau'}\sigma^{z}_{eg}+J^{ef}_{\tau \tau'}\sigma_{fe}^z\right) \\
&-\left(J^{ge}_{\tau \tau'}-J^{ef}_{jk}\right)\left(a_{\tau}a_{\tau'}\sigma^{fg}+a^\dagger_{\tau}a^\dagger_{\tau'}\sigma^{gf}\right)\bigg]\\
\label{equ:SW_TotalHamiltonian}
\end{aligned}
\end{equation}
where $\tilde{\omega}_e = \omega_e-\sum_\tau\chi^{ge}_\tau$, $\tilde{\omega}_f = \omega_f-\sum_\tau\chi^{ef}_\tau$, and
\begin{equation}
J^{jk}_{\tau \tau'}=\frac{1}{2}g_\tau g_{\tau'}\left(\frac{1}{\omega_\tau+\omega_j-\omega_k}+\frac{1}{\omega_{\tau'}+\omega_j-\omega_k}\right)
\end{equation}
and 
\begin{equation}
\Delta^{jk}_\tau = \omega_\tau + \omega_j- \omega_k\quad \quad \chi_\tau^{jk} = \frac{g_\tau^2}{\Delta^{jk}_\tau }    
\end{equation}
( with $\tau$th cavity frequency $\omega_\tau$ and three level transmon energy levels indices $j$ and $k$)

For a maximally anharmonic transmon with $\omega_f \rightarrow \infty$ this leads to a separation of the bare transmon level energy scales. If we focus on the low energy sector of $\ket{g/e}$, we arrive at the two-level description of \eqref{equ:TwoCavDispHam} from the main text.

\section{\label{app:SNAPGate}Eigen-SNAP gate}
Here we derive the effect of the transmon drive. For this we take the total Hamiltonian from Eq.~\eqref{equ:SW_TotalHamiltonian} and disregard the fast rotating terms $\propto (a_\tau a_{\tau'})^{(\dagger)}$. We call the resulting Hamiltonian $\tilde{H}$. Transforming into the frame rotating with $\tilde{H}$
\begin{equation}
    \tilde{H}_I(t)= e^{it\tilde{H}}(\tilde{H}+H_D) e^{-it\tilde{H}}
\end{equation}
and assuming the cavities are tuned into resonance, the time evolution to zeroth order in the Magnus expansion valid for $T_g\chi \gg 1$ is given by
\begin{equation}
\begin{aligned}
     U(t) &=\exp\left(-i\int_0^t\tilde{H}_I(t')dt'\right)\\
     &=\exp\left(\left(\hat{\eta}^{ge}(t)\sigma^{ge}  + \hat{\eta}^{ef}(t)\sigma^{ef}\right)- \text{h.c.}\right)\\
\end{aligned}  
\label{equ:UnitaryEvolution}
\end{equation}
with 
\begin{equation}
\begin{aligned}
\hat{\eta}^{ge}(t) 
     &=  -i \int_0^t \varepsilon (t')e^{it'2\left(2\chi^{ge}-\chi^{ef}\right)b_+^\dagger b_+} \\
\hat{\eta}^{ef}(t) 
     &=  -i \int_0^t \varepsilon (t')e^{it'2\left(2\chi^{ef}-\chi^{ge}\right)b_+^\dagger b_+} \\
\end{aligned}
\end{equation}
We emphasize, that the drive preserves the occupation number of the eigenmode $b_+$, which we made use of in the optimization scheme.
Since we want to address the $\sigma^{ge}$ transition, we drive on resonance with the occupation number-dependent frequencies using the driving scheme
\begin{equation}
\begin{aligned}
        \varepsilon(t) = \sum_n \varepsilon_n e^{i\omega_{eg}t}
        e^{-it2\left(2\chi^{ge}-\chi^{ef}\right)n}e^{i\Theta_n(t)} \quad .
\end{aligned}
\label{equ:DrivingQubitParametrization3Level}
\end{equation}
The resulting time evolution becomes simple in the two transmon level limit $\omega_f \rightarrow \infty$. Then using the pulse sequence described in  Sec.~$\ref{subsec:eigen_snap_gate}$:
\begin{equation}
\begin{aligned}
 U\left(\frac{T_g}{2},0\right)\ket{g}\ket{n}_+\ket{m}_- &=e^{-i\frac{\pi}{2}}\sin(\frac{T_g}{2}\abs{\varepsilon_n})\ket{e}\\
 &\quad \otimes \ket{n}_+\ket{m}_-\\
 U\left(T,\frac{T_g}{2}\right)\ket{e}\ket{n}_+\ket{m}_- &=e^{i\left(\theta +\frac{\pi}{2}\right)}\sin(\frac{T_g}{2}\abs{\varepsilon_n})\ket{g}\\
 &\quad \otimes \ket{n}_+\ket{m}_-\\
    \end{aligned}
\end{equation}
Hence, choosing the amplitude such that $T_g \varepsilon_n = \pi$, the time evolution operator evaluates to the $S^+$ gate
\begin{equation}
    S^+_n(\theta) = U\left(T,0\right)= e^{i\theta\ketbra{n}_+ }
\end{equation}
We can map this action on a general initial state $\ket{g}\ket{n}_+\ket{m}_-$ back to the non-interaction picture at times $\pi/\chi^{ge}$ that only rotates with the bare energy of the cavities $\omega_{c}$ because
\begin{equation}
\begin{aligned}
  e^{-i\tilde{H}\pi/\chi^{ge}} \ket{g}\ket{n,m}_c 
  &= e^{-i\pi/\chi^{ge}(\omega_g+\omega_c (a_1^\dagger a_1 +a_0^\dagger a_0 ))}\ket{g}\ket{n,m}_c  \quad . 
\end{aligned}
\end{equation}

\section{\label{app:EigenModeMapping}Eigenmode to cavity mode mapping}
In the following, we elaborate on the decomposition of logical codewords into their eigenmode occupation number state contributions. For this we discuss a central relation between the two-cavity Fock states $\ket{n,m}_c$ and $\ket{m,n}_c$ written in the eigenmode basis.

Given single qubit codewords $\ket{0/1}_\text{logic}$ in the cavity fock states $\ket{n}_c$
\begin{equation}
        \begin{aligned}
        \ket{0/1}_\text{logic} &=\sum_n c_n^{(0)/(1)} \ket{n}_c\\
    \end{aligned}
    \label{equ:SinglCavCode}
\end{equation}
the general two-qubit codewords can be written as

\begin{equation}
        \begin{aligned}
        \ket{00}_\text{logic} =&\;\sum_{n>m} C_n^{(0)}C_m^{(0)} \left(\ket{n,m}_c + \ket{m,n}_c\right) \\&\;+ \sum_{n}  \left(C_n^{(0)}\right)^2 \ket{n,n}_c \\
        \ket{01}_\text{logic} =&\;\sum_{n,m} C_n^{(0)}C_m^{(1)}\ket{n,m}_c  \\
        \ket{10}_\text{logic} =&\;\sum_{n,m} C_n^{(0)}C_m^{(1)}\ket{m,n}_c  \\
        \ket{11}_\text{logic} =&\;\sum_{n>m} C_n^{(1)}C_m^{(1)} \left(\ket{n,m}_c + \ket{m,n}_c\right) \\&\;+ \sum_{n}  \left(C_n^{(0)}\right)^2 \ket{n,n}_c \;.\\
    \end{aligned}
\label{equ:GenCodeWordTwoCav}
\end{equation}
The two cavity Fock states can be expressed through the eigenmodes $b_+$ and $b_-$
\begin{equation}
\begin{aligned}
    a_1 = \frac{1}{\sqrt{2}}\left(b_++b_-\right) \quad\quad a_0 = \frac{1}{\sqrt{2}}\left(b_+-b_-\right) \; .
\end{aligned}
\end{equation}
Then a general state in the cavity $\ket{n,m}_c = \ket{n}_{c_1}\otimes\ket{n}_{c_0}$ can be expressed by
\begin{equation}
\begin{aligned}
    \ket{n,m}_c =& \; \frac{1}{\sqrt{2^{n+m}\cdot n!m!}}\sum_j^n\sum_k^m\binom{n}{j}\binom{m}{k}(-1)^{j}\\
    &\cdot \left(b_+^\dagger\right)^{n+m-j-k}\left(b_-^\dagger\right)^{j+k}\ket{0}_+\ket{0}_-\\
=& \; \sum_j^n\sum_k^m c_{jk}^{nm}\;\text{sgn}_{jk}^{nm}\ket{n+m-j-k}_+\ket{j+k}_-
\end{aligned}
\label{equ:CavInEigenbasis}
\end{equation}
where $\text{sgn}_{jk}^{nm}$ is the sign of the $(j,k)$ contribution to the $\ket{n,m}_c $ state. Then one finds that the magnitude under the exchange of cavity occupation numbers is the same, $c_{jk}^{nm} = c_{jk}^{mn}$, but the sign flips depending on the parity of $j+k$ (Table~\ref{tab:sign_relationships}).
\begin{table}[h]
\centering
\renewcommand{\arraystretch}{1.5} 
\begin{tabular}{|c|c|}
\hline
\(j+k\) is even & \(\text{sgn}_{jk}^{nm} = \text{sgn}_{jk}^{mn}\) \\
\hline
\(j+k\) is odd & \(\text{sgn}_{jk}^{nm} = -\text{sgn}_{jk}^{mn}\) \\
\hline
\end{tabular}
\caption{Sign relationships of the eigenmode occupation number states under exchange of cavity occupation numbers depending on the parity of \(l+k\).}
\label{tab:sign_relationships}
\end{table}
Hence, if $n+m$ is even (odd), all odd (even) eigenmode occupation number states of the $b_+$ mode $\ket{\cdot}_+$ of the cavity states $\ket{n,m}_c$ and $\ket{m,n}_c$ will have opposite signs. Comparing with the general form of a two-cavity codeword in Eq.~\eqref{equ:GenCodeWordTwoCav} this proves the decomposition of the eigenmode decomposition into the logical codeword Eq.~\eqref{equ:EvenOdd} and Eq.~\eqref{equ:EvenOdd_OppositeParity}.

\section{\label{app:Optimized}Optimized parametrization}

\begin{figure}[t]
    \centering
    \includegraphics[width=\columnwidth]{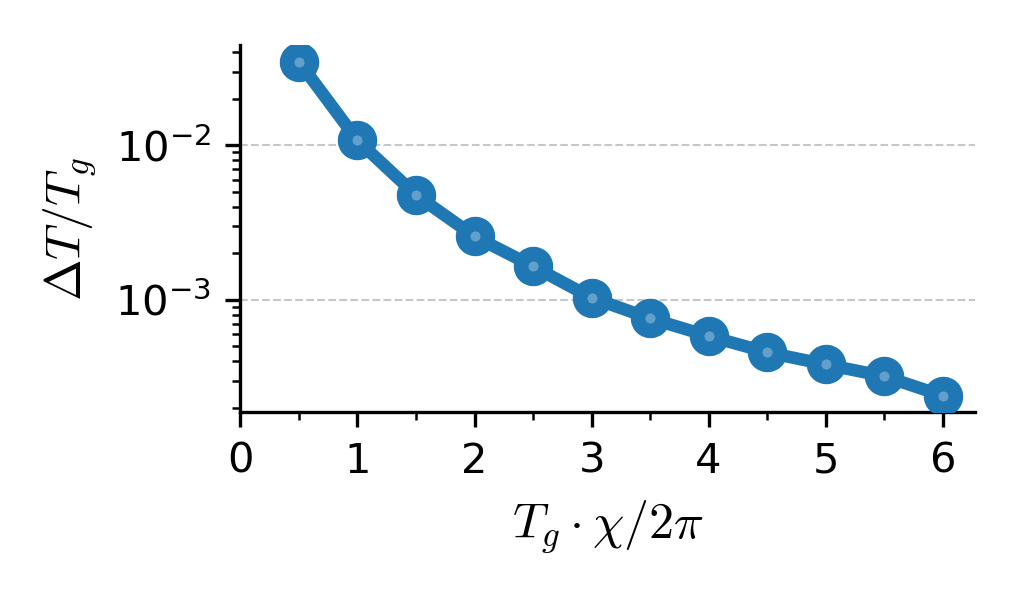} 
    \caption{Detuning of the total gate time for different total initial gate times }
    \label{fig:ThetaDetunigin}
\end{figure}

\begin{figure*}[t]
    \centering
    \includegraphics[width=\textwidth]{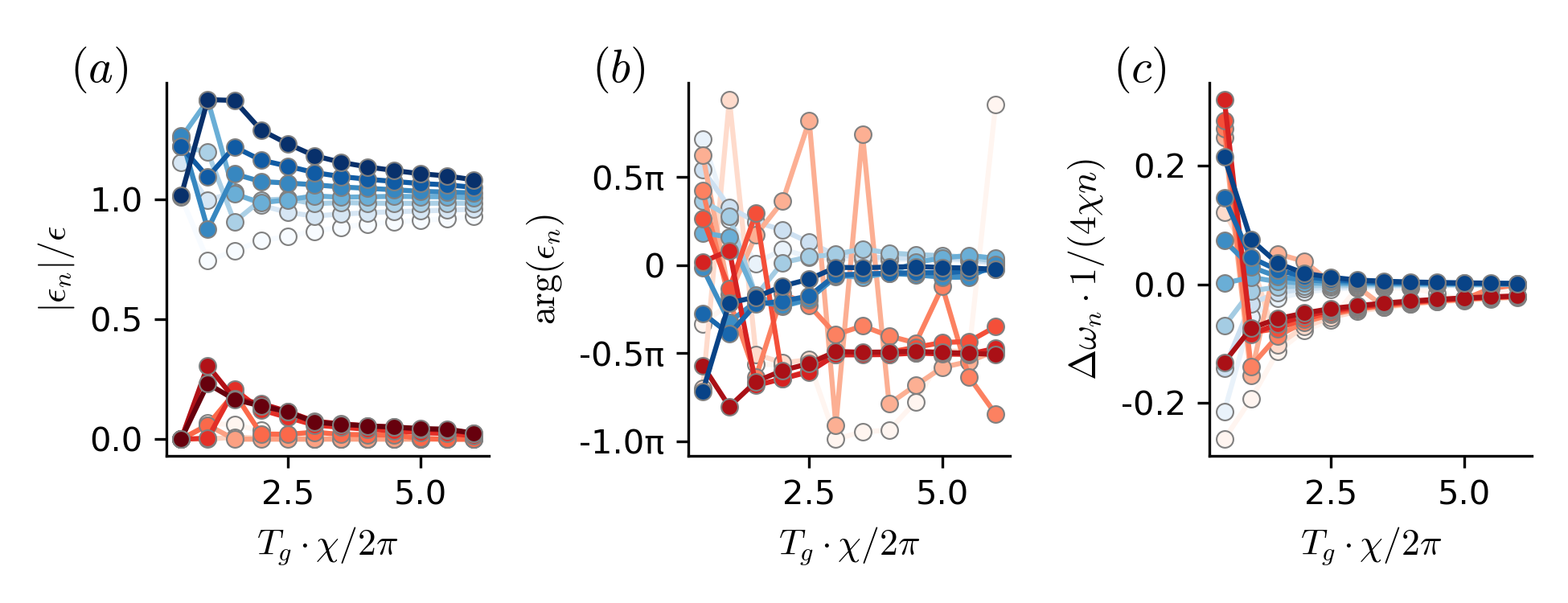}
    \caption{Optimized parameters of the drive (Eq.~\eqref{equ:DrivingQubitParametrization}) for the occupation number states at finite gate times. The odd (even) occupation number states in the range  $0-13$ are given by the blue (red) data points. The darker the color, the larger the $n$.  (a) The optimized amplitude $|\varepsilon_n|$ of the $n$th drive relative to the initial amplitude given by the Ansatz $\epsilon \cdot T_g = \pi$. For large times, the optimized amplitudes for the even and odd number drives approach the Ansatz. (b) The optimized phase of the odd number states goes to zero, again approaching the Ansatz, while the even component quieres a phase of $\approx \pi/2$ even at large gate times. (c) The detuning of the drive pulses diverges for short gate times for number states $n$ but approaches zero for long gate times.}
    \label{fig:ThetaOptimized}

\end{figure*}

The optimized parametrization is shown graphically for the total gate time detuning in Fig.~\ref{fig:ThetaDetunigin} and for the driving amplitude, phase, and frequency detuning in Fig.~\ref{fig:ThetaOptimized} respectively.\\
The time evolution of the system with the full master equation is computationally expensive. Therefore, in the optimization, we choose to only evolve with the effective Hamiltonian found in Monte Carlo jump methods, without performing the actual jumps:
\begin{equation}
    H_\text{eff} = H_\text{disp} - \frac{i}{2}\sum_j L^\dagger_j L_j \quad.
\end{equation}
The idea is that minimizing the non-Hermitian evolution and hence the jump probability allows the optimizer to take into account if a trajectory is highly susceptible to a loss process, without having to evolve the full master equation. Further, the computation is simplified by propagating in the frame rotating with the bare frequencies of the cavities and transmon, $e^{iH_0 t}H_\text{eff}e^{-iH_0 t}$ with $H_0$ given in Equations~\eqref{equ:TwoCavDispHamParts}.
In the optimization and evaluation we simulate the system using the adaptive step-size Runge-Kutta method (RK45).
In the optimization we used the ADAM-optimizer to adjust the update step size. We optimized up to convergence of the fidelity. For short gate times this meant about $3000$ optimization steps. For long gate times, the optimization converges faster, since the Ansatz parametrization is already closer to the optimal parametrization.

\section{\label{app:Fidelity}Minimal fidelity estimate\protect\\ }
We evaluate the fidelity of the logical quantum gate numerically by calculating
\begin{equation}
\begin{aligned}
        F &= \frac{1}{16} \sum_n \tr \left[ \sigma^{ij}_\text{logic}\mathcal{E}\left(\sigma^{ij}_\text{logic}\right)\right]\\
        \sigma^{ij}_\text{logic}&= U_\text{logic} \left(U_\text{target}\sigma_i \otimes \sigma_jU^\dagger_\text{target}\right) U_\text{logic}^\dagger
\end{aligned}
\label{equ:FidelityExpression}
\end{equation}
where $\mathcal{E}$ is quantum channel implemented by the time evolution, $U_\text{target}$ is the target logical gate, $\sigma_j$ with $j=0,1,2,3$ are the Pauli matrices and $U_\text{logic}$ is the logical codeword encoding matrix $U_\text{logic} = \{ \ket{00}_\text{logic}, \ket{01}_\text{logic}, \ket{10}_\text{logic}, \ket{11}_\text{logic}\}$. 

We determine the minimal infidelity of the binomial encoding (Eq.~\eqref{equ:BinomCode}) by expanding the density matrix in powers of the decoherence strength $\gamma T_g$ 
\begin{equation}
    \rho(t) = \rho^{(0)}(t) + \rho^{(1)}(t) +  \rho^{(2)}(t) + \cdots
\end{equation}
and calculating the time evolution up to first order using the Lindblad master equation
\begin{equation}
    \frac{d\rho}{dt} = -i [H_\text{disp}+H_D, \rho] + \mathcal{D}\left[\rho(t)\right]
\label{equ:LindbladMasterEquation}
\end{equation}
with
\begin{equation}
    \mathcal{D}\left[\rho(t)\right] = \gamma \left( L \rho L^\dagger - \frac{1}{2} \{ L^\dagger L, \rho \} \right) \quad .
\end{equation}
This treatment is analogous to the procedure used in \cite{landgrafFastQuantumControl2023} but adopted to the two-cavity system. Further, the evaluation of the fidelity in Eq.~\ref{equ:FidelityExpression} is done to align with the $\text{even}/\text{even}$ decomposition of the eigenmode occupation number states used in the main text.

We assume the loss contributions of the cavity relaxation, transmon decay, and dephasing to be additive, such that they can be treated individually.
Without the loss of generality, we focus on the stroboscopic times for which the gate time $T_g$ is given by a multiple of $\pi/\chi^{ge}$. Then zeroth order evolution in the limit of $T_g\chi\gg1$ is simply given by unitary $U(t)$ of Eq.~\eqref{equ:UnitaryEvolution}
\begin{equation}
    \rho^{(k)}(t) = U(t) \rho(0)  U^\dagger(t)
\end{equation}
The corrections to this ideal trajectory are given by the higher order contributions, $k>0$, with 
\begin{equation}
    \rho^{(k)}(t) = U(t) \left(\int_{0}^{t} U^\dagger(s) \mathcal{D}\left[\rho^{(k-1)}(s)\right] U(s) \, ds \right)  U^\dagger(t) \quad .
\end{equation}
For the amplitude damping in the cavity with Lindblad operators
\begin{equation}
\begin{aligned}
    L_0 &= \sqrt{\kappa}a_0 = \sqrt{\frac{\kappa}{2}}\left(b_+- b_-\right) \\
    L_1 &=\sqrt{\kappa} a_1 = \sqrt{\frac{\kappa}{2}}\left(b_+ + b_-\right) \quad
\end{aligned}
\end{equation}
and a binomial encoding protecting against amplitude loss, the fidelity to  first order is given by
\begin{equation}
    F(\rho^{(0)} + \rho^{(1)} ) = 1-\kappa T_g\, \operatorname{Tr}\left[ \rho^{(0)} \left( b^\dagger_+ b_+ + b^\dagger_- b_- \right) \right]
\end{equation}
which simply measures the average occupation number of the encoding in the two cavities. Here we have used the fact, that the term $ b_- \rho^{(0)}b_-^\dagger$ leads to a state outside of the logical subsapce. It therefore does not contribute to the fidelity. By construction, a binomial encoding has the same occupation number for any $\rho$, which is element of the logical subspace. For the encoding in Eq.~\eqref{equ:BinomCode}, the mean total cavity occupation number is $\langle n_1 + n_0\rangle=6$.  

For the transmon dephasing and transmon amplitude damping, the evaluation of the first-order contribution to the fidelity is more intricate. Different from the cavity photon loss, the loss in the transmon depends on the initial state of the cavity, since the trajectory in the transmon subspace depends on the parity of the two-cavity eigen-mode occupation number. To stay in the language of even/odd contributions, we divide the the total initial density matrix $\rho = \rho_c \otimes \ketbra{g}$ into even/odd contributions, since the time evolution operator acts differently onto these two contributions. Then the fidelity takes the form
\begin{equation}
\begin{aligned}
        F(\rho) =&\; \Tr \left[\rho_\text{target}^\dagger\mathcal{E}(\rho)\right]\\
=&\;\sum_{\tau,\tau'\in\text{even/odd}}\Tr \left[\ketbra{g}\mathcal{E}_{\tau,\tau'}\left(\ketbra{g}\right)\right]\\
&\quad\quad\quad\quad\;\;\cdot\Tr \left[\rho_c\left(P_{\tau}\rho_cP_{\tau'}\right)\right]\\
\end{aligned}
\end{equation}
where $P_{\tau}$ projects onto the eigen-mode occupation number parity $\tau \in \text{even/odd}$. The quantum process $\mathcal{E}_{\tau,\tau'}$ indicates, that in the transmon subspace, the ket is propagated according to parity $\tau$, while the bra is propagated according to the parity $\tau'$. Evaluating this expression to first order, we find for $L_-=\sqrt{\gamma_-}\sigma^{ge}$ 
\begin{equation}
\begin{aligned}
&\mathcal{E}_{\text{even},\text{even}} = 
\begin{pmatrix}
0 & 0 \\
0 & 0
\end{pmatrix}; \quad
\mathcal{E}_{\text{even},\text{odd}} = T_g
\begin{pmatrix}
\frac{i}{4} & \frac{1 + i}{4\pi} \\
0 & 0
\end{pmatrix};\\
&\mathcal{E}_{\text{odd},\text{even}} = T_g
\begin{pmatrix}
-\frac{i}{4} & 0 \\
\frac{1+ i}{4\pi} & 0
\end{pmatrix}; \quad
\mathcal{E}_{\text{odd},\text{odd}} = T_g
\begin{pmatrix}
-\frac{3}{8} & \frac{1 + i}{2\pi} \\
\frac{1 - i}{2\pi} & \frac{3}{8}
\end{pmatrix}
\end{aligned}
\end{equation}
and for $L_z=\sqrt{\gamma_z}\sigma^z_{eg} $
\begin{equation}
\begin{aligned}
&\mathcal{E}_{\text{even},\text{even}} = T_g
\begin{pmatrix}
0 & 0 \\
0 & 0
\end{pmatrix}; \quad
\mathcal{E}_{\text{even},\text{odd}} = T_g
\begin{pmatrix}
i & \frac{1 + i}{\pi} \\
0 & 0
\end{pmatrix};\\
&\mathcal{E}_{\text{odd},\text{even}} = T_g
\begin{pmatrix}
-i & 0 \\
\frac{1 + i}{\pi} & 0
\end{pmatrix}; \quad
\mathcal{E}_{\text{odd},\text{odd}} = T_g
\begin{pmatrix}
-\frac{1}{2} & 0 \\
0 & \frac{1}{2}
\end{pmatrix}.
\end{aligned}
\end{equation}
The resulting minimal fidelity evaluated over the complete logical basis is given by Eq.~\eqref{equ:FidTheoryMin}.

\section{\label{app:Diag}Hamiltonian Representation}
In the following we show the equivalence between the Hamiltonian in the eigenmode $b_\pm$ and the cavity mode $a_{1/2}$ representation, given by Eq.~\eqref{equ:TwoCavDispHamEigen} and \eqref{equ:TwoCavDispHamParts} respectively. Here we assume cavities in resonance, $J_{01}^{ge} = \chi$. Substituting Eq.~\eqref{equ:eigenmodes} into the first row of Eq.~\eqref{equ:TwoCavDispHamEigen} 
\begin{equation}
\begin{aligned}
    &\left(\frac{1}{2}\omega_{eg}-2\chi b^\dagger_+b_+\right)\sigma^{z}_{eg}\\
    &= \frac{1}{2}\omega_{eg}\sigma^{z}_{eg} - 2\chi \frac{1}{2}\left(a^\dagger_1 + a^\dagger_0\right)\left(a_1 +  a_0\right)\sigma^{z}_{eg}\\
    &= \frac{1}{2}\omega_{eg}\sigma^{z}_{eg}- \chi \sigma^{z}_{eg} \left(a^\dagger_1 a_1 + a^\dagger_0 a_0 +  a^\dagger_1 a_0 + a^\dagger_0a_1\right) \\
    &=\frac{1}{2}\omega_{eg}\sigma^{z}_{eg}- \chi \sigma^{z}_{eg} \sum_{\tau = 0, 1}a^\dagger_\tau a_\tau - \chi \sigma^{z}_{eg} \left( a^\dagger_1 a_0 + a^\dagger_0a_1\right) 
\end{aligned}
\end{equation}
one recovers the terms of the dispersive coupling $H_\chi$, qubit dependent coupling $H_J$ and bare qubit energy of Eq.~\eqref{equ:TwoCavDispHamParts}. Substituting the eigenmode definition into the second row of Eq.~\eqref{equ:TwoCavDispHamEigen} 
\begin{equation}
\begin{aligned}
    &\omega_c\left(b^\dagger_+b_+ +b^\dagger_-b_-  \right)\\
    &= \omega_c\left(\frac{1}{2}\left(a^\dagger_1 + a^\dagger_0\right)\left(a_1 +  a_0\right) +\frac{1}{2}\left(a^\dagger_1 - a^\dagger_0\right)\left(a_1 - a_0\right) \right)\\
    &=  \omega_c\bigg(\frac{1}{2}\left(a^\dagger_1 a_1+ a^\dagger_0a_0 + a^\dagger_1 a_0 + a^\dagger_0a_1\right)\\
    &\quad\quad\;+\frac{1}{2}\left(a^\dagger_1 a_1+ a^\dagger_0a_0 - a^\dagger_1 a_0 - a^\dagger_0a_1\right) \bigg)\\
    &= \omega_c \left( a^\dagger_1 a_1 + a^\dagger_0 a_0\right)
\end{aligned}
\end{equation}
we recover the bare energy of the cavities and therefore the full $H_0$ of Eq.~\eqref{equ:TwoCavDispHamParts}.

\end{document}